\documentclass[prl,aps,preprintnumbers,twocolumn,floatfix]{revtex4}
\usepackage{epsfig}
\usepackage{amsmath}
\usepackage{graphicx,xcolor}

\newcommand{\vect}[1]{\boldsymbol{#1}}
\def\pt{\vect{p}_T}
\def\Kt{\vect{K}_T}
\def\ptmod{{p}_T}
\def\Ktmod{{K}_T}
\def\Phperp{\vect{P}_{h\perp}}
\def\PT{{P}_{h\perp}}
\def\pup{P^{\uparrow}}

\begin{document}

\preprint{JLAB-THY-11-1466}
\preprint{NIKHEF-2011-033}
\preprint{YITP-SB-11-48}

\title{Calculation of Transverse-Momentum-Dependent Evolution for Sivers Transverse Single Spin Asymmetry Measurements}

\author{S.~Mert~Aybat$^{1}$, Alexei~Prokudin$^{2}$ and Ted~C.~Rogers$^{3}$}
\affiliation{$^{1}$Nikhef Theory Group, Science Park 105, 1098XG Amsterdam, The Netherlands}
\affiliation{${}^2$Jefferson Lab, 12000 Jefferson Avenue, Newport News, Virginia 23606, USA}
\affiliation{${}^3$C.N.\ Yang Institute for Theoretical Physics, 
Stony Brook University, Stony Brook, New York 11794--3840, USA}
\date{May 29, 2012}

\begin{abstract}
The Sivers transverse single spin asymmetry (TSSA) 
is calculated and compared at different scales using the TMD evolution equations applied to previously 
existing extractions.  
We apply the Collins-Soper-Sterman (CSS) formalism, 
using the version recently developed by Collins.   
Our calculations rely on the universality properties of TMD-functions that follow from the TMD-factorization theorem.  
Accordingly, the non-perturbative input is fixed by earlier 
experimental measurements, including both polarized semi-inclusive deep inelastic scattering (SIDIS)
and unpolarized Drell-Yan (DY) scattering.  
It is shown that recent preliminary COMPASS measurements are consistent with the suppression prescribed by TMD evolution. 
\end{abstract}

\maketitle

Conventional collinear perturbative QCD (pQCD), when applied in its range 
of applicability, has proven 
successful for over three 
decades~\cite{Kronfeld:2010bx}.  Along the way, it has illustrated the importance of evolution for relating physical 
observables to 
fundamental quark-gluon QCD degrees of freedom in a unified formalism.
The classic applications of pQCD require parametrizations of 
collinear parton distribution functions (PDFs), wherein all intrinsic transverse motion of the confined partons is neglected inside the 
hard part of the collision and integrated over in the definitions of the PDFs. 
The PDFs contain information about the intrinsic non-perturbative structure of the hadron 
and have clear operator definitions with 
well-understood scale dependence (QCD evolution).  

An important next step is to achieve a similarly successful application of perturbative QCD that 
takes into account the emerging picture of the hadron as a three-dimensional 
dynamical object composed of partons with their own intrinsic motion.  In such studies, the relevant 
observables are not 
properly handled by standard collinear factorization, and one is confronted with the objects like    
TMD parton distribution functions (TMD PDFs)  and TMD fragmentation functions (FFs).  
(Collectively, we refer to such objects as ``TMDs")
These, like their collinear 
counterparts, describe the non-perturbative properties of the external hadrons; but, 
unlike the collinear PDFs, the TMDs also account for the intrinsic transverse Fermi motion
of the bound 
partons.
 
Several experimental facilities, including HERMES (DESY), COMPASS (CERN) 
and JLab,  explore these distributions.
Moreover, the future 12 GeV JLab upgrade and a planned Electron Ion Collider (EIC)~\cite{Boer:2011fh} will provide new opportunities to 
experimentally probe hadron structure.
A particularly interesting TMD PDF is the so-called Sivers function, which is interpreted as the probability density for finding a parton with a given 
transverse and longitudinal momentum inside a transversely polarized hadron (usually proton) target.  That it 
arises at leading power in pQCD is due to 
interesting 
non-perturbative aspects of QCD related to time reversal and parity invariance.  
In polarized SIDIS, it gives a $\sin(\phi_h-\phi_S)$ azimuthal 
modulation to the differential cross section, $\phi_S$ and $\phi_h$ being the azimuthal angles of the 
initial transverse hadron spin and the final state hadron transverse momentum respectively.  This letter
will focus on a comparison of recent theoretical treatments of the Sivers function with recently available experimental data on 
TSSAs in SIDIS, and predict the size of the asymmetry for future extractions at larger $Q$.

Like the collinear PDFs, TMDs evolve with the hard scale $Q$.  
To properly account for this it is imperative to work in a
 QCD factorization 
formalism that incorporates well-defined TMDs.
The original TMD-factorization formalism  was developed by Collins, Soper, and Sterman~\cite{Collins:1981uk,Collins:1981uw,Collins:1984kg} in the context of 
$e^+ e^-$-annihilation to back-to-back jets and the unpolarized Drell-Yan process.  The CSS formalism was later extended by Ji, Ma and Yuan to 
SIDIS in Ref.~\cite{Ji:2004wu}, and to include polarization in Ref.~\cite{Ji:2004xq}.  
The application of Collins-Soper (CS) evolution was extended to the spin-dependent case by Idilbi et al in Ref.~\cite{Idilbi:2004vb}.
Furthermore, an  
implementation of a CS-style evolution has been applied 
in Ref.~\cite{Boer:2001he} to the calculation of TSSAs by 
including a Sudakov form factor, leading to 
approximate power-law $Q$-behavior for the 
peak of the
asymmetry.

For quite some time, however, a satisfactory treatment of TMD-factorization remained  
incomplete 
because of
a lack of 
good
definitions for the TMD PDFs and FFs themselves~\cite{Collins:2003fm,Collins:2008ht,Hautmann:2009zzb,Cherednikov:2008uk}. 
The most commonly quoted definitions suffered from 
unphysical divergences, rendering it 
unclear which objects should be used in parametrizations of experimental data and treated as universal in the usual sense of a factorization theorem.  
This is particularly problematic for studies that purport to extract properties intrinsic to the hadrons.
Recently, a 
complete TMD-factorization derivation, 
in terms of well-defined TMDs with individual 
evolution properties, was presented by Collins in Ref.~\cite{collins}.  
Refs.~\cite{Aybat:2011zv,Aybat:2011ge} further illustrated how the formalism 
can be used to obtain evolved TMDs from fixed-scale fits for unpolarized TMDs and the Sivers function.

Existing extractions of the Sivers function using the TSSA, $A_{UT}^{\sin(\phi_h-\phi_S)}$, have been performed 
using experimental data at fixed scales 
 ~\cite{Efremov:2004tp,Vogelsang:2005cs,Anselmino:2005nn,Arnold:2008ap,Anselmino:2008sga,Anselmino:2011gs}. 
 These extractions provide interesting information about TMD effects at the fixed scales where they are performed; 
 however, without a reliable way to evolve them to different scales, their predictive power remains limited.  

The purpose of this letter is to demonstrate that by using QCD evolved TMDs
one can explain an observed discrepancy between HERMES and COMPASS data and 
for the first
time make
predictions for upcoming experiments at higher energy scales on the basis of a complete and correct treatment of evolution for the TMDs.

{\bf Definitions and Notation:} The differential cross section for SIDIS, $l(l) + N(P,S)\to l(l^{\prime}) + h(P_h) + X$ is \cite{Mulders:1995dh,Diehl:2005pc,Bacchetta:2006tn}
 \begin{equation}
 \label{sidis}
 \frac{d\sigma}{dxdydzd\phi_hd\phi_S \PT d\PT}  = \frac{\alpha^2 y}{2 z Q^4}\,M\,  L_{\mu\nu} W^{\mu\nu}\,
 \end{equation}
 where $\PT$ is the transverse momentum of the final state hadron $h$, and where 
we utilize the standard kinematical variables: $q^2 = -Q^2$, $x={Q^2}/{2P\cdot q}$, $y={P \cdot q}/{P\cdot l}$,$z={P\cdot P_h}/{P\cdot q}$.
The TMD-factorization formula for SIDIS in terms of well-defined TMD PDFs is~\cite{collins}
\begin{multline}
\label{hadron_tensor} 
W^{\mu\nu} =  \sum_{f}|{\cal H}_f(Q^2,\mu)|^{\mu\nu}  \\ \times \int d^2\pt d^2\Kt \delta^{(2)}(z \pt + \Kt - \Phperp)  \\ \times F_{f/\pup}(x,z\ptmod,S;\mu,\zeta_F) 
 D_{h/f}(z,\Ktmod;\mu,\zeta_D)  \\
+ Y(\PT,Q) \;,
\end{multline}
where all non-perturbative information is encoded in the TMD PDF $F_{f/\pup}$ and the TMD FF $D_{h/f}$ while $|{\cal H}_f(Q^2,\mu)|^{\mu\nu}$ is a perturbatively calculable hard part.
The $Y(\PT,Q)$-term gives the correct treatment of the cross section at high $\PT \sim Q$ in terms of collinear factorization.  
As is common, the renormalization scale is set to $\mu = Q$.  The parameters $\zeta_F,\zeta_D$, which are related to the regularization of rapidity divergences, obey $\zeta_F \zeta_D \sim Q^4$.  
(Consistency with perturbation theory requires $\mu$, $\sqrt{\zeta_F}$ and $\sqrt{\zeta_D}$ each to be of order $\mathcal{O}(Q)$.)

The Sivers asymmetry is defined as a the ratio of cross section combinations:
 \begin{multline}
 \label{sivers_asymmetry}
A_{UT}^{\sin(\phi_h-\phi_S)}  = \\  \frac{\int d\phi_h d\phi_s 2 \sin(\phi_h-\phi_S) (\sigma(\phi_h,\phi_S) - \sigma(\phi_h,\phi_S+\pi))}
{\int d\phi_h d\phi_S (\sigma(\phi_h,\phi_S) + \sigma(\phi_h,\phi_S+\pi))}.
 \end{multline}
In the numerator, the integration over azimuthal angles with 
a $\sin(\phi_h-\phi_S)$ weighting factor projects out the Sivers effect.
The numerator and denominator may also be integrated over $x$, $z$ and/or $\PT$ depending on the particular
combination of variables one is interested in.  

The asymmetry is obtained by applying the TMD-factorization in Eq.~\eqref{hadron_tensor} to obtain cross sections in Eq.~\eqref{sivers_asymmetry}. 
The calculations themselves are typically done in transverse coordinate $b_T$-space in terms of structure functions, whose 
relations to the differential cross section are given in Ref.~\cite{Boer:2011xd}.
In the case of the Sivers function, 
the general expression for 
the evolved TMD in coordinate space was found in Ref.~\cite{Aybat:2011ge} to be
\begin{eqnarray}
    \tilde{F}^{\prime \, \perp \, f}_{1T}(x,b_T;Q,\zeta_F) 
&=&    \tilde{F}^{\prime \, \perp \, f}_{1T}(x,b_T;Q_0,Q_0^2)  \nonumber\\ 
&\ &\hspace{-30mm}\times
\exp \Biggl\{ 
         \ln \frac{Q}{Q_0} \tilde{K}(b_{\ast};\mu_b)
         + \int_{Q_0}^Q \frac{d \mu^\prime}{\mu^\prime} \Big[ \gamma_F(g(\mu^\prime);1) \nonumber\\
&\ &\hspace{-10mm}
                 - \ln \frac{Q}{\mu^\prime} \gamma_K(g(\mu^\prime)) \Big] 
\nonumber\\
&\ &\hspace{-30mm}
         + \int_{Q_0}^{\mu_b} \frac{d \mu^\prime}{\mu^\prime} \ln \frac{Q}{Q_0}
                       \gamma_K(g(\mu^\prime))
         - g_K(b_T) \ln \frac{Q}{Q_0}
\Biggr\}.
\label{eq:evolvedsiv1}
\end{eqnarray}
Analogous formulas hold for the unpolarized TMDs.
The symbols, $\gamma_K$ and $\gamma_F$, are perturbatively calculable anomalous dimensions, $\tilde{K}(b_{\ast};\mu_b)$ is 
the perturbatively calculable CS kernel written in terms of  $b_{\ast}$ which is the 
prescription for matching to the region where $1/b_T$ can be treated as a perturbatively large scale.
We use the usual prescription of~\cite{Collins:1984kg} where 
$b_{\ast} = b_T/\sqrt{1+b_T^2/b_{max}^2}$ and $\mu_b = C_1/b_{\ast} $, and $b_{max}$ and $C_1$ 
are parameters to be specified later.
Note that it is the {\it derivative} $\tilde{F}_{1T}^{\prime q\perp}(x,b_T;Q,\zeta)$ of the QCD 
evolved coordinate-space Sivers function with respect to $b_T$ that appears
in Eq.~\eqref{eq:evolvedsiv1} for the evolution.  $Q_0$ is the starting scale for the evolution.
The non-perturbative but universal and scale-independent 
function $g_K(b_T)$ describes the behavior of $\tilde{K}(b_{T};\mu_b)$ in the non perturbative region at large $b_T$.  
An important prediction from the TMD-factorization theorem is that $g_K(b_T)$ is universal, 
not only between different processes, but also 
between all different 
types 
of 
quark TMDs (both PDFs and FFs). 
\begin{figure*}[t]
\centering
  \begin{tabular}{c@{\hspace*{5mm}}c}
    \includegraphics[scale=0.4]{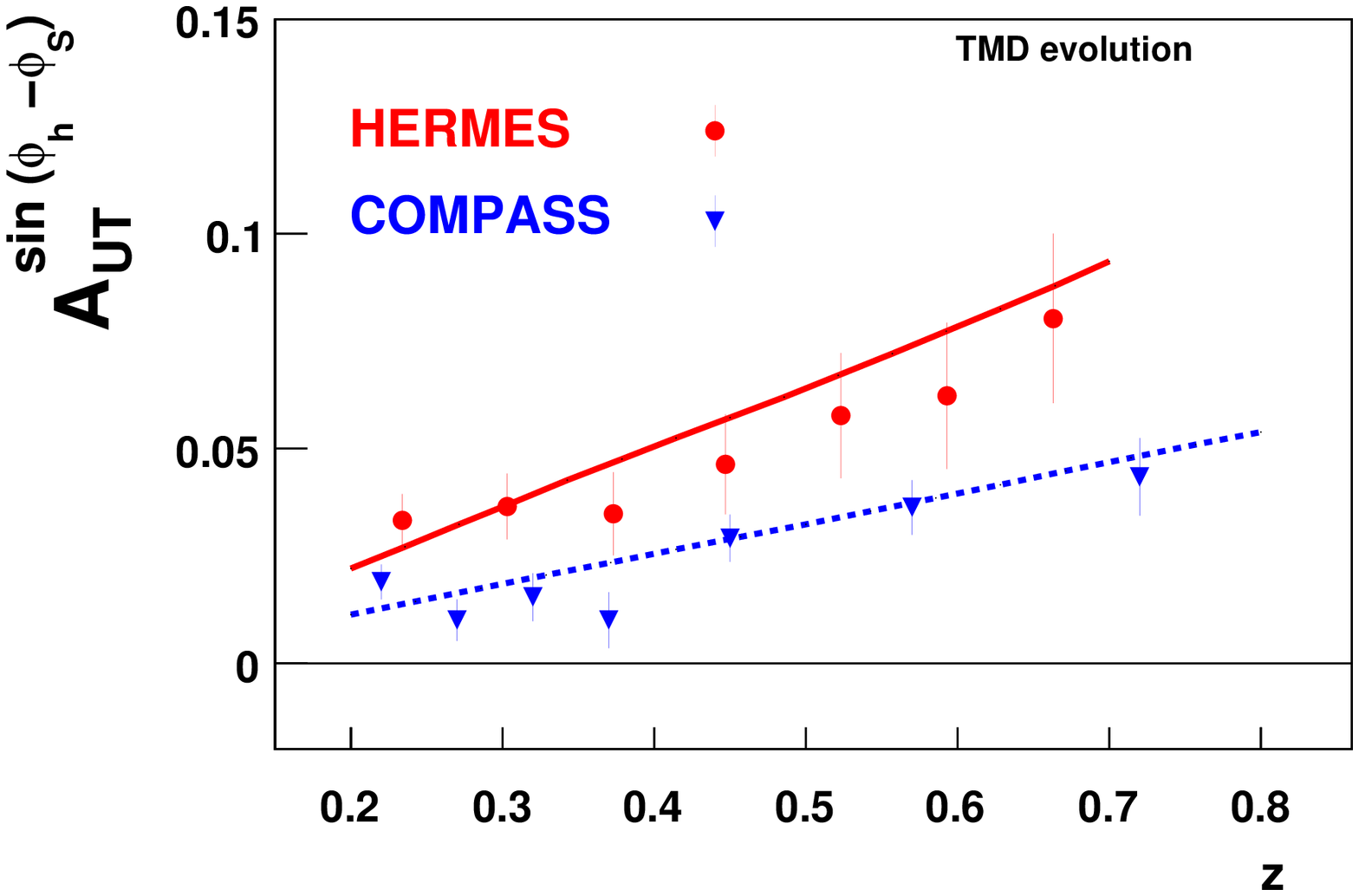}
    &
    \includegraphics[scale=0.4]{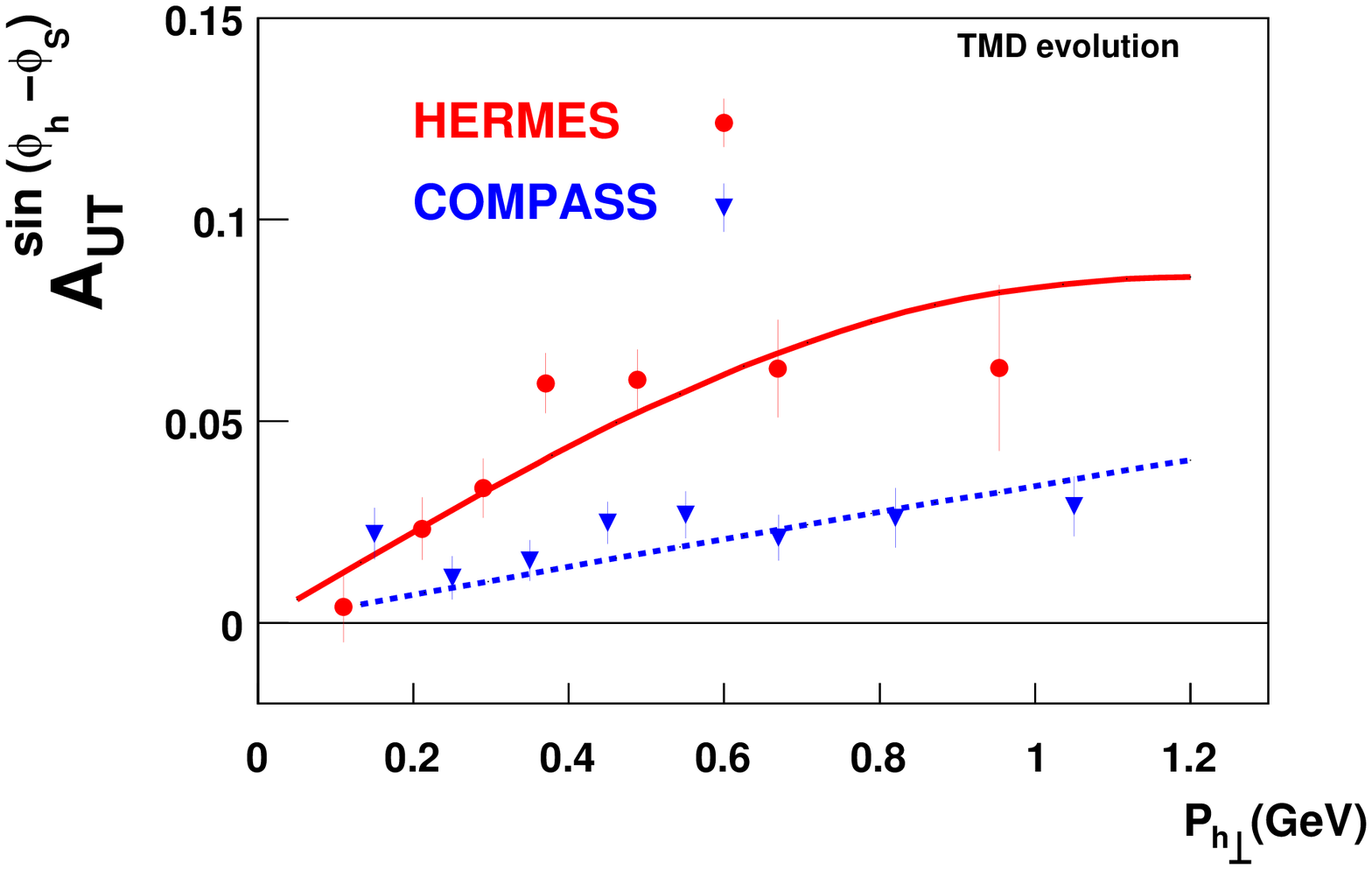}
  \\
  (a) & (b)
  \end{tabular}
\caption{Comparison between HERMES \cite{Collaboration:2009ti} and preliminary COMPASS data \cite{Bradamante:2011xu} for the (a.) $z$ and (b.) $\PT$ dependence of
Eq.~\eqref{sivers_asymmetry} with a proton target 
and $\pi^+$ and $h^+$ as final state hadrons respectively. 
The solid line is the fit from Ref.~\cite{Anselmino:2011gs}.  
The dashed curve is the result of evolving to the COMPASS scale using the full TMD-evolution of Ref.~\cite{Aybat:2011ge}.}
\label{fig:evolvedgraphs}
\end{figure*}

For this letter, we assume that 
$Q$ is low enough that we can neglect the $Y$-term in Eq.~\eqref{hadron_tensor}~\cite{Aybat:2011zv}.
Furthermore, we use a Gaussian ansatz to parametrize the input distribution $\tilde{F}^{\prime \, \perp \, f}_{1T}(x,b_T;Q_0,Q_0^2)$, though 
this means that we do not utilize the fact that at larger $\PT \gg \Lambda_{\rm QCD}$ the TMD PDFs are related to collinear distributions 
through 
perturbative coefficient functions.  
(In the Sivers case, 
this
involves the Qiu-Sterman function~\cite{Qiu:1998ia,Kang:2011mr}.)
Still, in Ref.~\cite{Aybat:2011zv} it was shown that a Gaussian ansatz provides a good description of the 
evolved Sivers function for the low region of transverse momentum and 
moderate 
hard scales we are interested in for this letter.
Several groups have parametrized the polarized and unpolarized TMD PDFs and FFs at fixed scales
in terms of simple Gaussian fits \cite{Efremov:2004tp,Anselmino:2005nn,Arnold:2008ap,Anselmino:2008sga,Anselmino:2011gs,Bacchetta:2011gx}, 
and these may be used as the input functions for the evolution.

{\bf Analysis and Discussion:} 
As input distributions, we use the 
already existing Gaussian parametrizations of the 
Torino group~\cite{Anselmino:2011gs}, relevant for low $\langle Q^2 \rangle_{\rm Hermes}\simeq 2.4$ GeV$^2$ and typical for the HERMES experiment.  
These earlier fixed scale fits were done at leading order in QCD
and neglecting the QCD evolution of the TMDs, which was not available at that time. 
We note that the analysis of Ref.~\cite{Anselmino:2011gs} also uses deuteron data \cite{Alekseev:2008dn} from the COMPASS experiment, which 
corresponds to higher values of
$Q^2$. However, the COMPASS asymmetry \cite{Alekseev:2008dn} on 
the
deuteron target is very small due to strong cancellations 
between the up and down quark Sivers functions and thus is not heavily affected by the evolution. 
We have verified that the results of the Torino fits
are negligibly altered if the deuterium data are excluded and only HERMES data \cite{Collaboration:2009ti} are used in the fit, and the main result of our present 
analysis is not affected. 
 
Our calculations will follow the steps of Ref.~\cite{Aybat:2011ge}.
For $g_K$, we use the functional 
form $g_K = \frac{1}{2} g_2 b_T^2$ with $g_2 = 0.68$ GeV$^2$ 
~\cite{Landry:2002ix}, which was obtained by fits performed using Drell-Yan data. In 
Eq.~\eqref{eq:evolvedsiv1}, this corresponds to using $C_1 = 1.123$ and $b_{max} = 0.5$ GeV$^{-1}$. All anomalous dimensions and $\tilde K$ are 
calculated to lowest non-vanishing order as in Refs.~\cite{collins,Aybat:2011zv}.  

In Fig.~\ref{fig:evolvedgraphs}(a,b), we show the evolution using the full 
TMD-factorization approach as expressed in Eq.~\eqref{eq:evolvedsiv1}, where the evolution  
is due to
the terms in the exponential.
The evolution is applied to the most recent Torino fits~\cite{Anselmino:2011gs} as a function $z$ and $\PT$ , and 
use hard scales corresponding to both HERMES data \cite{Collaboration:2009ti} and recent preliminary COMPASS data \cite{Bradamante:2011xu}.  The result of the evolution is compared with the data.
The $x$-dependent asymmetry is not ideal 
for the comparison because there are strong correlations between $x$ and $Q^2$. (Recall $Q^2\simeq x y s$.) 
However, $z$ or $\PT$ dependent asymmetries are measured at almost the same hard scales, 
namely $\langle Q^2\rangle_{\rm Hermes}\simeq 2.4$ GeV$^2$ 
and $\langle Q^2 \rangle_{\rm COMPASS}\simeq 3.8$ GeV$^2$, so we focus on the Sivers asymmetry as a function of these variables.  
(For the preliminary $h^+$ COMPASS data that we use, $\langle Q^2 \rangle$ varies between $3.63$~GeV$^2$ and $3.88$~GeV$^2$, in the range of $z$ from $0.2$ to $0.7$.
The corresponding variation in our calculation is negligible relative to the variation between the HERMES and preliminary COMPASS data sets.)
We observe that including QCD evolution leads to excellent consistency between the HERMES \cite{Collaboration:2009ti}  and preliminary 
COMPASS data \cite{Bradamante:2011xu}, without the need for further fitting.  
The two data sets correspond to different ranges in $x$, and this could be partly responsible for the variation.  
A similarly fast evolution has not been seen so far in the Collins Single Spin Asymmetry~\cite{Bradamante:2011xu,Airapetian:2010ds}, suggesting a more complicated 
interplay between $b_T$, $x$ and $z$ dependence.  We leave a careful consideration of these issues to future studies. 
Nevertheless, we
find the early success of the comparison in Fig.~\eqref{fig:evolvedgraphs} encouraging,
especially as leading order fits \cite{Anselmino:2005nn,Anselmino:2008sga,Anselmino:2011gs} fail to reproduce
COMPASS proton data \cite{Bradamante:2011xu} sufficiently well.
Still, we caution that future fits will need to account for the 
$x$-dependence as well.

A critical point is that the information about the non-perturbative evolution contained in $g_K$ is taken from the measurement of 
a totally different observable, at much higher energy scales~\cite{Landry:2002ix} (unpolarized Drell-Yan scattering up to Tevatron energies).  
In Fig.~~\ref{fig:evolvedgraphs}(b) we show a similar plot but for the $\PT$ dependence.
That the same 
$g_K$ successfully describes TSSA at HERMES and COMPASS is compelling evidence for the 
universality of $g_K$ predicted by the TMD-factorization theorem.

\begin{figure}
	\includegraphics[width=0.5\textwidth]{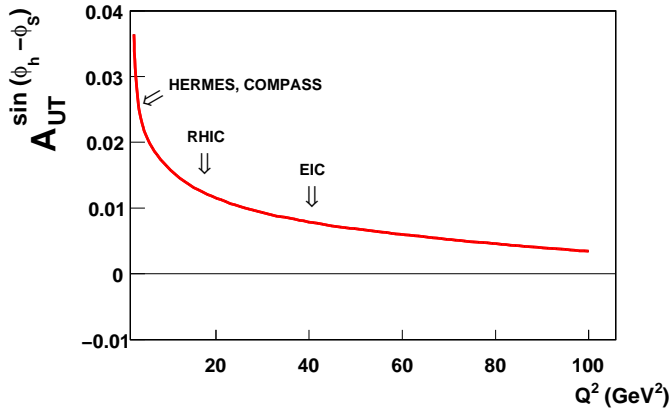}
	\caption{ Sivers evolution in $Q^2$, integrated over $x$, $z$ and $\PT$.\label{fig:fig4}}
\end{figure}

In Fig.~\ref{fig:fig4}, we show the evolution of the full asymmetry to higher values of $Q^2$.
Note that, although Refs.~\cite{Aybat:2011ge,Aybat:2011zv} 
report a strong suppression of the unpolarized 
TMDs and the Sivers function itself with increasing $Q^2$, the TSSA is not as heavily 
suppressed.  Therefore, it may be expected that the Sivers SSA remains significant at the higher $Q$ values of experiments planned 
at the Relativistic Heavy Ion Collider (RHIC) and the EIC.
Still, the QCD evolution effects are clearly non-negligible and should be correctly included in future extractions.  
Ref.~\cite{Boer:2001he} predicts a roughly $\sim 1 / \sqrt{Q}$ suppression for the {\it peak} of the 
Sivers asymmetry as a function of transverse momentum, for large $Q^2 \gtrsim 10$~GeV$^2$.
We find that, for the full asymmetry integrated over all transverse momentum, a power-like scaling law does not provide a good description in the range of $Q$ in Fig.~\ref{fig:fig4}.
Generally, we find that the evolution leads to suppression 
that is faster than $\sim 1/\sqrt{Q}$,  but slower than $\sim 1/Q^2$. 
We caution, however, that 
a completely correct treatment at large $Q$ must include the $Y$-term in Eq.~\eqref{hadron_tensor}, and it is possible that this
will weaken the rate of the suppression.  

To conclude, we remark that it is important for future theoretical calculations to
not only explain experimental results, but also to make precise pQCD-based predictions that can be tested against future data at larger $Q$.  
With this in mind, we view the success of the TMD-factorization treatment in explaining the HERMES and COMPASS as  highly encouraging.

\acknowledgments
We are very grateful to M.~Anselmino, C.~Aidala, D.~Boer, E. Boglione, J.~Collins, E.~Laenen, S. Melis, M.~Pennington,  J.~W.~Qiu, and G.~Sterman for helpful discussions.
M.~Aybat and T.~Rogers thank Jefferson Lab for kind hospitality.
Authored by a Jefferson Science Associate, LLC under U.S. DOE Contract 
No. DE-AC05-06OR23177.  
T.~Rogers was supported by the National Science Foundation, grant PHY-0969739. M.~Aybat acknowledges 
support from the research program of the ``Stichting voor Fundamenteel Onderzoek der Materie (FOM)'', which is 
financially supported by the ``Nederlandse Organisatie voor Wetenschappelijk Onderzoek (NWO)''.

\bibliography{ssa}

\end{document}